\documentclass[runningheads,a4paper]{llncs}

\usepackage{amssymb}
\usepackage{graphicx}

\usepackage[cmex10]{amsmath}
\usepackage{subfig}

\usepackage{url}
\newcommand{\keywords}[1]{\par\addvspace\baselineskip
\noindent\keywordname\enspace\ignorespaces#1}

\begin{document}

\mainmatter

\title{Model-Independent Analytic Nonlinear \\ Blind Source Separation}

\titlerunning{Model-Independent Nonlinear Blind Source Separation}

\author{David N. Levin%
%\thanks{}%
}

\authorrunning{Model-Independent Nonlinear Blind Source Separation}

\institute{Dept. of Radiology, University of Chicago,\\
1310 N. Ritchie Ct., Unit 26 AD, Chicago, IL 60610\\
Email: d-levin@uchicago.edu\\
\url{http://radiology.uchicago.edu/directory/david-n-levin}}

\toctitle{}
\tocauthor{}
\maketitle

\begin{abstract}
Consider a time series of measurements of the state of an evolving system, $x(t)$, where $x$ has two or more components. This paper shows how to perform nonlinear blind source separation; i.e., how to determine if these signals are equal to linear or nonlinear mixtures of the state variables of two or more statistically independent subsystems. First, the local distributions of measurement velocities are processed in order to derive vectors at each point in $x \mbox{-space}$. If the data are separable, each of these vectors must be directed along a subspace of $x \mbox{-space}$ that is traversed by varying the state variable of one subsystem, while all other subsystems are kept constant. Because of this property, these vectors can be used to construct a small set of mappings, which must contain the ``unmixing'' function, if it exists. Therefore, nonlinear blind source separation can be performed by examining the separability of the data after it has been transformed by each of these mappings. The method is analytic, constructive, and model-independent. It is illustrated by blindly recovering the separate utterances of two speakers from nonlinear combinations of their audio waveforms.
\end{abstract}

\section{Introduction}
\label{introduction}

The signals from a process of interest are often contaminated by signals from extraneous processes, which are thought to be statistically independent of the process of interest but are otherwise unknown. This raises the question: can one use the observed signals to determine if two or more independent processes are present, and, if so, can one derive a representation of the evolution of each of them? In other words, if a system is effectively evolving in a closed box, can one process the signals emanating from the box in order to learn the number and nature of the subsystems within it? There is a variety of methods for solving this blind source separation (BSS) problem for the special case in which the signals are linearly related to the underlying independent subsystem states (\cite{Ans}, \cite{Comon Jutten}). However, some observed signals (e.g., from biological or economic systems) may be nonlinear functions of the underlying system states. Computational methods of separating such nonlinear mixtures are limited (\cite{Jutten}, \cite{Almeida}), even though humans seem to do it in an effortless manner.

Consider an evolving physical system that is being observed by making time-dependent measurements ($x_{k}(t) \mbox{ for } k = 1, \ldots ,N$ where $N \geq 2$), which are coordinates on the system's state space. In Conclusion, we describe how to choose measurements that comprise such coordinates. The objective of blind source separation is to determine if the measurement time series is separable; i.e., to determine if it can be transformed into another coordinate system, $s$ (called the ``source'' or ``separable'' coordinate system), in which the transformed time series describes the evolution of statistically independent subsystems. Specifically, we want to know if there is an invertible, possibly nonlinear, $N \mbox{-component}$ ``unmixing'' function, $f$, that transforms the measurement time series into a source time series:
\begin{equation}
\label{mixture}
s(t) = f[x(t)] ,
\end{equation}
where the $N$ components of $s(t)$ can be partitioned into statistically independent, possibly multidimensional groups.

This paper utilizes a criterion for ``statistical independence'' ~\cite{Levin-bss-JAP} that differs from the conventional one. Specifically, let $\rho_S(s,\dot{s})$ be the probability density function (PDF) in $(s,\dot{s}) \mbox{-space}$, where $\dot{s}~=~ds/dt$. Namely, let $\rho_S(s,\dot{s}) ds d\dot{s}$ be the fraction of total time that the location and velocity of $s(t)$ are within the volume element $ds d\dot{s}$ at location $(s,\dot{s})$. In this paper, the data are defined to be separable if and only if there is an unmixing function that transforms the measurements so that $\rho_S(s,\dot{s})$ is the product of the density functions of individual components (or groups of components)
\begin{equation}
\label{phase space factorization}
\rho_S(s,\dot{s}) = \prod_{a=1,2, \ldots}{\rho_{Sa}(s_{(a)},\dot{s}_{(a)}}) .
\end{equation}
where $s_{(a)}$ is a subsystem state variable, comprised of one or more of the components of $s$. This criterion for separability is consistent with our intuition that the statistical distribution of the state and velocity of any independent subsystem should not depend on the particular state and velocity of any other independent subsystem.

This criterion for statistical independence should be compared to the conventional criterion, which is formulated in $s \mbox{-space}$ (i.e., state space) instead of $(s,\dot{s}) \mbox{-space}$ (the space of states and state velocities). In particular, let $\rho_S(s)$ be the PDF, defined so that $\rho_S(s) ds$ is the fraction of total time that the trajectory $s(t)$ is located within the volume element $ds$ at location $s$. In some formulations of the BSS problem, the system is said to be separable if and only if there is an unmixing function that transforms the measurements so that $\rho_S(s)$ is the product of the density functions of individual components (or groups of components)
\begin{equation}
\label{state space factorization}
\rho_S(s) = \prod_{a=1,2, \ldots}{\rho_{Sa}(s_{(a)})} ,
\end{equation}
In \emph{every} formulation of BSS, multiple solutions can be created by applying ``subsystem-wise'' transformations, which transform each subsystem's components among themselves. These solutions are the same as one another, except for differing choices of the coordinate systems used to describe each subsystem. However, the criterion in (\ref{state space factorization}) is so weak that it suffers from a much worse non-uniqueness problem: namely, solutions can almost always be created by mixing the state variables of \textit{different} subsystems of other solutions (see \cite{Hosseini}, \cite{Jutten}, \cite{Hyvarinen-uniqueness}).

There are at least two reasons why (\ref{phase space factorization}) is the preferred way of defining ``statistical independence'':
\begin{enumerate}
\item If a physical system is comprised of two independent subsystems, we normally expect that there is a unique way of identifying the subsystems. As mentioned above, (\ref{state space factorization}) is too weak to meet this expectation. On the other hand, (\ref{phase space factorization}) is a much stronger constraint than (\ref{state space factorization}). Specifically, (\ref{state space factorization}) can be recovered by integrating both sides of (\ref{phase space factorization}) with respect to velocity. This shows that the solutions of (\ref{phase space factorization}) are a subset of the solutions of (\ref{state space factorization}). Therefore, it is certainly possible that (\ref{phase space factorization}) reformulates the BSS problem so that it has a unique solution (up to subsystem-wise transformations), although this is not proved in this paper.
\item For all systems that obey the laws of classical physics and are in thermal equilibrium at temperature $T$, the PDF in $(s,\dot{s}) \mbox{-space}$ is proportional to the Maxwell-Boltzmann distribution~\cite{Reif}
\begin{equation}
\label{M-B distribution}
e^{-E(s,\dot{s})/(kT)}
\end{equation}
where $E$ is the system's energy and $k$ is the Boltzmann constant. If the system consists of two non-interacting subsystems, the system's energy is the sum of the subsystem energies
\begin{equation}
\label{summed energies}
E = E_{1}(s_{(1)},\dot{s}_{(1)}) + E_{2}(s_{(2)},\dot{s}_{(2)})
\end{equation}
where $s_{(1)}$ and $s_{(2)}$ are subsystem state variables comprised of one or more  components of $s$.
This demonstrates that, for all classical systems composed of non-interacting subsystems, the system's PDF in $(s,\dot{s}) \mbox{-space}$ is the product of the subsystem PDFs in $(s,\dot{s}) \mbox{-space}$, as stated in (\ref{phase space factorization}).
\end{enumerate}

There are several other ways in which the proposed method of nonlinear BSS differs from methods in the literature:
\begin{enumerate}
\item As stated above, in this paper the BSS problem is reformulated in the joint space of states and state velocities. Although there is some earlier work in which BSS is performed with the aid of velocity information (\cite{Ehsandoust}, \cite{Lagrange}), these papers utilize the \textit{global} distribution of measurement velocities (i.e., the distribution of velocities at all points in state space). In contrast, the method proposed here exploits additional information that is present in the \textit{local} distributions of measurement velocities (i.e., the velocity distributions in each neighborhood of state space).
\item Many investigators have attempted to simplify the BSS problem by assuming prior knowledge of the nature of the mixing function; i.e., they have modelled the mixing function. For example, the mixing function has been assumed to have parametric forms that describe post-nonlinear mixtures \cite{Taleb}, linear-quadratic mixtures \cite{Merrikh}, and other combinations (\cite{Duarte}, \cite{Yang}, \cite{Tan}). In contrast, the present paper proposes a model-independent method that can be used in the presence of any invertible diffeomorphic mixing function.
\item In many other approaches, nonlinear BSS is reduced to the optimization problem of finding the unmixing function that maximizes the independence of the source signals corresponding to the observed mixtures. This usually requires the use of iterative algorithms with attendant issues of convergence and computational cost (e.g., \cite{Comon Jutten}, \cite{Duarte}). In contrast, the method proposed in this paper is analytic and constructive. Specifically, the observed data are used to construct a small collection of mappings, $\{u(x)\}$, that must contain an unmixing function, if one exists. To perform BSS, it then suffices to determine if any of these functions transforms the measured time series, $x(t)$, into a time series, $u[x(t)]$, having a factorizable PDF. The data are separable if and only if this is the case.
\end{enumerate}

There are two earlier papers (\cite{Levin-bss-JAP}, \cite{Levin-IEEE-Trans}) that utilize the criterion in (\ref{phase space factorization}) in order to perform nonlinear BSS. However, both of these approaches are quite different from the one proposed here. The current paper shows how the measurement time series endows state space with local vectors that contain crucial information about the separability of the data. Specifically, if the data are separable, each of these vectors must be directed along a subspace of $x \mbox{-space}$ that is traversed by varying the state variable of one subsystem, while all other subsystem variables are kept constant. Because of this property, these vectors can be used to determine if the data are separable and, if they are, to determine the transformation to a separable coordinate system. In contrast, the presence of these vectors played no role whatsoever in the methods discussed in \cite{Levin-IEEE-Trans} and \cite{Levin-bss-JAP}. Instead,
\begin{enumerate}
\item In \cite{Levin-IEEE-Trans}, BSS was performed by deriving a large number of local scalars that must lie in low-dimensional subspaces, if the data were separable. The vectorial structure on state space was not utilized or even recognized.
\item Likewise, local vectors also played no role in \cite{Levin-bss-JAP}. Instead, the local second-order velocity correlation matrix was taken to define a Riemannian metric on the space of measurements ($x$). Then, nonlinear BSS was performed by using differential geometry to look for a transformation to another coordinate system ($s$), in which this metric was block-diagonal everywhere.
\end{enumerate} 
In short, although \cite{Levin-bss-JAP}, \cite{Levin-IEEE-Trans}, and the present paper all utilize the same criterion for statistical independence (i.e., (\ref{phase space factorization})), these three approaches differ greatly in how they determine whether this criterion is satisfied by a given signal.

\begin{figure*}[!tbp]
\begin{center}
\includegraphics[trim=4cm 0 0 0,width=6.5in]{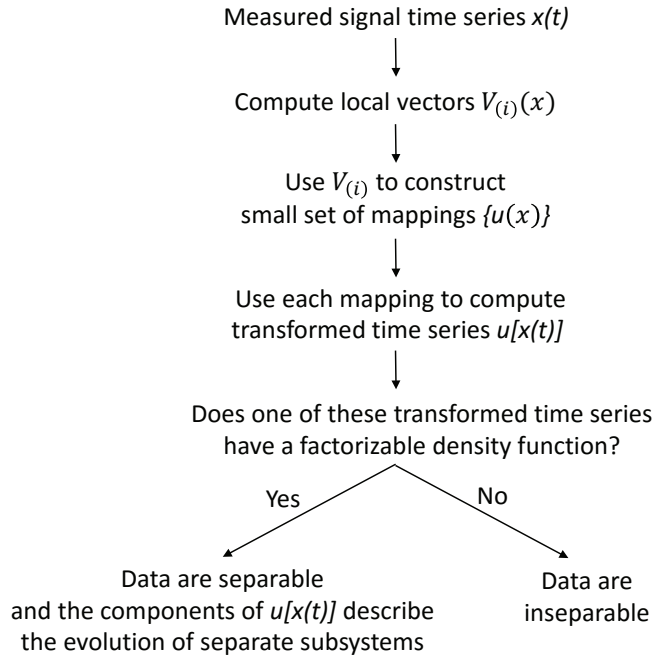}
\caption{ The proposed method of nonlinear blind source separation}
\label{figure1}
\end{center}
\end{figure*}

The next section gives a detailed description of the proposed method of nonlinear blind  source separation, which is schematically illustrated in Figure~\ref{figure1}. Section~\ref{experiments} illustrates the method by using it to blindly recover the utterances of two speakers from nonlinear mixtures of their audio waveforms.  The last section discusses the implications of this approach. Note that brief versions of this work were presented in ~\cite{Levin arXiv}, ~\cite{Levin ITISE}, and ~\cite{Levin LCA-ICA}.

\section{Method}
\label{method}

For didactic purposes, the next subsection describes a five-step procedure for performing nonlinear BSS of systems with two degrees of freedom. Then, Subsection \ref{N dimensional systems} describes how to generalize this procedure so that it can be applied to systems and subsystems having any number of degrees of freedom.

\subsection{Systems having two degrees of freedom}
\label{two-dimensional systems}

\textit{1. The local second- and fourth-order correlations of the measurement velocity ($\dot{x}$) are computed in small neighborhoods of the measurement space. These correlations are used to compute two local vectors ($V_{(i)}(x) \mbox{ for } i = 1,2$)}\\
 The first step is to construct second-order and fourth-order local correlations of the data's velocity
\begin{equation}
\label{C2 definition}
C_{kl}(x) = \, \langle (\dot{x}_k-\bar{\dot{x}}_k) (\dot{x}_l-\bar{\dot{x}}_l) \rangle_{x} 
\end{equation}
\begin{equation}
\label{C4 definition}
\begin{split}
C_{klmn}(x) = \, \langle (\dot{x}_k-\bar{\dot{x}}_k) & (\dot{x}_l-\bar{\dot{x}}_l) \\ 
& (\dot{x}_m-\bar{\dot{x}}_m) (\dot{x}_n-\bar{\dot{x}}_n) \rangle_{x}
\end{split}
\end{equation}
where $\bar{\dot{x}} = \langle \dot{x} \rangle_x$, where the bracket denotes the time average over the trajectory's segments in a small neighborhood of $x$, and where all subscripts are integers equal to $1$ or $2$. Because $\dot{x}$ is a contravariant vector, $C_{kl}(x)$ and $C_{klmn}(x)$ are local contravariant tensors of rank $2$ and $4$, respectively. The definition of the PDF implies that $C_{kl}(x)$ and $C_{klmn}(x)$ are two of its moments; e.g.,
\begin{equation}
\label{PDF moment}
C_{kl \ldots}(x) = \frac {\int \rho(x,\dot{x}) (\dot{x}_k-\bar{\dot{x}}_k) (\dot{x}_l-\bar{\dot{x}}_l) \ldots d\dot{x}} {\int \rho(x,\dot{x}) d\dot{x}} ,
\end{equation}
where $\rho(x,\dot{x})$ is the PDF in the $x$ coordinate system, where ``$\ldots$'' denotes possible additional subscripts on the left side and corresponding additional factors of $\dot{x}-\bar{\dot{x}}$ on the right side, and where all subscripts are integers equal to $1$ or $2$. Although (\ref{PDF moment}) is useful in a formal sense, in practical applications all required correlation functions can be computed directly from local time averages of the data (e.g., (\ref{C2 definition})-(\ref{C4 definition})), without explicitly computing the data's PDF. Also, note that velocity ``correlations'' with a single subscript vanish identically
\begin{equation}
\label{C_k=0}
C_k(x)=0  .
\end{equation}

Next, let $M(x)$ be any local $2 \times 2$ matrix, and use it to define $M \mbox{-transformed}$ velocity correlations, $I_{kl}$ and $I_{klmn}$
\begin{equation}
\label{I2 definition}
I_{kl}(x) = \sum_{1 \leq k', \, l' \leq 2} M_{kk'}(x) M_{ll'}(x) C_{k'l'}(x) ,
\end{equation}
\begin{equation}
\label{I4 definition}
\begin{split}
I_{klmn}(x) = \sum_{1 \leq k', \, l', \, m', \, n' \leq 2} & M_{kk'}(x) M_{ll'}(x) \\
& M_{mm'}(x) M_{nn'}(x) C_{k'l'm'n'}(x) .
\end{split}
\end{equation}
Because $C_{kl}(x)$ is generically positive definite, it is possible to find a particular form of $M(x)$ that satisfies
\begin{equation}
\label{M definition 1}
I_{kl}(x) = \delta_{kl}
\end{equation}
\begin{equation}
\label{M definition 2}
\sum_{1 \leq m \leq 2} I_{klmm}(x) = D_{kl}(x) , 
\end{equation}
where $D(x)$ is a diagonal $2 \times 2$ matrix. Such an $M(x)$ can be constructed from the product of three matrices: 1) a rotation that diagonalizes $C_{kl}(x)$, 2) a diagonal rescaling matrix that transforms this diagonalized correlation into the identity matrix, 3) another rotation that diagonalizes
\begin{displaymath}
\sum_{1 \leq m \leq 2} \tilde{C}_{klmm}(x) ,
\end{displaymath} 
where $\tilde{C}_{klmn}(x)$ is the fourth-order velocity correlation ($C_{klmn}(x)$) after it has been transformed by the first rotation and the rescaling matrix.   As long as $D$ is not degenerate, $M(x)$ is unique, up to arbitrary \textit{local} permutations and/or reflections. In almost all applications of interest, the velocity correlations will be continuous functions of $x$. Therefore, in any neighborhood of state space, there will always be a continuous solution for $M(x)$, and this solution is unique, up to arbitrary \textit{global} permutations and/or reflections. 

In any other coordinate system $x'$, the most general solution for $M'$ is given by
\begin{equation}
\label{M'}
M'_{kl}(x') = \sum_{1 \leq m, \, n \leq 2} P_{km} M_{mn}(x) \frac{\partial x_n}{ \partial x'_l} ,
\end{equation}
where $M$ is a matrix that satisfies (\ref{M definition 1}) and (\ref{M definition 2}) in the $x$ coordinate system and where $P$ is a product of permutation and reflection matrices. This can be proven by substituting this equation into the definition of $I'_{kl}(x')$ and $I'_{klmn}(x')$ and by noting that these quantities satisfy (\ref{M definition 1}) and (\ref{M definition 2}) in the $x'$ coordinate system because (\ref{I2 definition})-(\ref{I4 definition}) satisfy them in the $x$ coordinate system. By construction, $M$ is not singular, and, therefore, it has a non-singular inverse.

Notice that (\ref{M'}) shows that the rows of $M$ transform as local covariant vectors, up to global permutations and/or reflections. Likewise, the same equation implies that the columns of $M^{-1}$ transform as local contravariant vectors (denoted as $V_{(i)}(x) \mbox{ for } i = 1, 2$), up to global permutations and/or reflections. As shown in the following, these particular vectors contain significant information about the separability of the data. In fact, they can be used to construct a mapping that must be an unmixing function, if one exists.\\

\textit{2. The $V_{(i)}(x)$ are used to construct a mapping, $u(x) = (u_{1}(x), u_{2}(x))$.}\\
Because we are considering systems with just two degrees of freedom, only one mapping $u(x)$ needs to be  constructed. However, to analyze systems with more than two degrees of freedom, a small set of mappings, $\{u(x)\}$, must be constructed, as suggested in Figure \ref{figure1} and as described in detail in Subsection \ref{N dimensional systems}. Working in the $x$ coordinate system, we begin by picking any point $x_0$. We then find a curve $X(\sigma)$ that passes through $x_0$ and is tangential to the local vector $V_{(1)}(x)$ at each point. Here, $\sigma$ denotes a variable that parameterizes the curve and increases monotonically as the curve is traversed in one direction. Formally, $X(\sigma)$ can be chosen to be a solution of the first-order differential equations
\begin{equation}
\label{X 1D}
\frac{dX}{d\sigma} = V_{(1)}(X)
\end{equation}
that satisfies the boundary condition, $X(0) = x_0$. Then, for each value of $\sigma$, we construct a curve, $Y(\tau)$, which passes through the point $X(\sigma)$ and is tangential to the local vector $V_{(2)}(x)$ at each point. Here, $\tau$ parameterizes this curve, increasing monotonically as it is traversed in one direction. Mathematically, $Y(\tau)$ can be chosen to be a solution of
\begin{equation}
\label{Y 2D}
\frac{dY}{d\tau} = V_{(2)}(Y)
\end{equation}
that satisfies the boundary condition, $Y(0) = X(\sigma)$. Finally, the function $u_{1}(x)$ is defined so that it is constant along each of the $Y$ curves. Specifically, $u_{1}(x) \equiv \sigma$ whenever $x$ is on the $Y$ curve passing through $X(\sigma)$. A function $u_{2}(x)$ can be defined by following an analogous procedure in which the roles of $V_{(1)}(x)$ and $V_{(2)}(x)$ are switched.\\

\textit{3. The mapping $u(x)$ is used to transform the measured time series, $x(t)$, into the time series $u[x(t)] = (u_{1}[x(t)],u_{2}[x(t)])$.}\\

\textit{4. It is determined if the components of $u[x(t)]$ are statistically independent.}\\
This can be done by computing its PDF and determining if it factorizes as
\begin{equation}
\label{1D u PDF factorization}
\rho_{U}(u,\dot{u}) = \prod_{a=1,2}{\rho_{Ua}(u_{a},\dot{u}_{a})} .
\end{equation}
Here, $u$ denotes $u[x(t)]$, and $\dot{u}$ is its time derivative. Alternatively, we can compute a large set of correlations of multiple components of $u[x(t)]$ and then determine if they are products of lower-order correlations, as required by (\ref{1D u PDF factorization}).\\

\textit{5. The result of step 4 is used to determine if the data are separable and, if they are, to determine an unmixing function. Specifically, if the components of $u[x(t)]$ are found to be statistically independent in step 4, it is obvious that the data are separable and $u(x)$ is an unmixing function. On the other hand, if the components of $u[x(t)]$ are found to be statistically dependent, the data are inseparable in any coordinate system.}\\
This last statement is a consequence of the following fact, which is proved in the next two paragraphs: namely, if the data are separable, the constructed mapping, $u(x)$, must be an unmixing function.

Before proving this, we show that the matrix $M$ and the $V_{(i)}(x)$ have simple forms in the separable coordinate system, $s$. In particular, we prove that the following diagonal matrix is the $M$ matrix in the $s$ coordinate system
\begin{equation}
\label{diagonal MS}
M_S(s) = \left( \begin{array}{ccc}  
  C^{-0.5}_{S11}(s_1) & 0 & \\
   0 & C^{-0.5}_{S22}(s_2) &
   \end{array} \right) ,
\end{equation}
where $C_{Skl}(s)$ for $k,l = 1, 2$ are the second-order velocity correlations in the $s$ coordinate system. This can be proved by demonstrating that $M_S$ satisfies (\ref{M definition 1}) and (\ref{M definition 2}) in the $s$ coordinate system. To do this, first note that (\ref{PDF moment}), (\ref{phase space factorization}), and (\ref{C_k=0}) imply that the second-order velocity correlations are diagonal in the $s$ coordinate system. It follows that (\ref{M definition 1}) is satisfied by $M_S$ in the $s$ coordinate system. Furthermore, it is not difficult to show that (\ref{M definition 2}) is also satisfied by $M_S$ in the $s$ coordinate system. To see this, substitute (\ref{diagonal MS}) into the sum on the left side of (\ref{M definition 2}) for $k \neq l$. Because of the diagonality of $M_S$, each term in this summation is proportional to a fourth-order velocity correlation in the $s$ coordinate system that has just one index equal to $1$ (or $2$) and the other three indices all equal to $2$ (or $1$). Each of these terms must vanish because of (\ref{PDF moment}), (\ref{phase space factorization}), and (\ref{C_k=0}). This completes the proof that $M_S$ satisfies both (\ref{M definition 1}) and (\ref{M definition 2}) in the $s$ coordinate system, and, therefore, it is the $M$ matrix in the $s$ coordinate system, as asserted above.

Because $M_S$ is diagonal, the local vectors in the $s$ coordinate system, denoted  $V_{S(1)}(s)$ and $V_{S(2)}(s)$, are oriented along the unit vectors, $(1,0)$ and $(0,1)$, respectively. Therefore, in the $s$ coordinate system, the curve, $X(\sigma)$, which was used in the definition of $u_{1}(x)$, is a horizontal straight line passing through the point $s[x_{0}]$, Similarly, each $Y$ curve is a vertical straight line passing through $s[X(\sigma)]$ for some value of $\sigma$. This implies that $s_{1}$ is constant along each $Y$ curve, being equal to the value of $s_1$ at its intersection with the $X$  curve. But, recall that $u_{1}(x)$ is also constant along each $Y$ curve, being equal to the value of $\sigma$ at its intersection with the $X$ curve. Therefore, because $\sigma$ is defined to vary monotonically along the $X$ curve and because the values of $s_1$ also vary monotonically along that curve, these paired values must be monotonically related to one another; i.e., $\sigma = h_{1}(s_{1})$ where $h_1$ is a monotonic function. It follows that $u_{1}(x)$ and $s_{1}(x)$ must also be monotonically related at each point; i.e., $u_{1}(x) = h_{1}[s_{1}(x)]$. In a similar manner, it can be shown that $u_{2}(x)$ and $s_{2}(x)$ are also related by some monotonic function. This means that $u_{1}(x)$ and $u_{2}(x)$ are component-wise transformations of $s_{1}(x)$ and $s_{2}(x)$. Because such component-wise transformations do not affect separability, it immediately follows that $u(x)$ is an unmixing function, as asserted above.

\subsection{Systems having any number of degrees of freedom}
\label{N dimensional systems}

This subsection describes how the procedure in Subsection \ref{two-dimensional systems} can be generalized to perform nonlinear BSS of systems having $N$ degrees of freedom, where $N \geq 2$. The overall strategy is to determine if the system can be separated into two (possibly multidimensional) independent subsystems. If the data cannot be so separated, they are simply inseparable. If such a two-fold separation is possible, the data describing the evolution of each independent subsystem can be examined in order to determine if it can be further separated into two lower-dimensional subsystems. This recursive process can be repeated until each independent subsystem cannot be further divided into lower-dimensional parts. For example, for $N = 3$, we can first determine if the system can be separated into a subsystem with one degree of freedom and a subsystem having two degrees of freedom. If such a separation is possible, the data describing the two-dimensional subsystem can then be examined to determine if it can be further subdivided into two one-dimensional subsystems.

The five-step procedure for performing nonlinear BSS is described below and illustrated in Figure \ref{figure1}.\\

\textit{1. The local second- and fourth-order correlations of the measurement velocity ($\dot{x}$) are computed in small neighborhoods of the measurement space. These correlations are used to compute $N$ local vectors ($V_{(i)}(x) \mbox{ for } i = 1, \ldots ,N$).}\\
This is done exactly as in step $1$ in Subsection \ref{two-dimensional systems}, except for the fact that: 1) each subscript can have any value between $1$ and $N$ (instead of $1$ and $2$); 2) each vector $V_{(i)}(x)$ has $N$ components (instead of two components).\\

\textit{2. The $V_{(i)}(x)$ are used to construct a small set of $N \mbox{-component}$ functions, $\{u(x)\}$, each of which is defined to be the union of two functions constructed with fewer components, $u_{(1)}(x)$ and $u_{(2)}(x)$.}\\
One such mapping is constructed for each way of partitioning the $V_{(i)}$ into two groups (groups $1$ and $2$), without distinguishing the order of the two groups or the order of vectors within each group. For example, for a three-dimensional system ($N = 3$), three $u(x)$ functions must be constructed, each one corresponding to one of the three distinct ways of partitioning three vectors into two groups: $\{\{V_{1}\}, \{V_{2}, V_{3}\}\}$, $\{\{V_{2}\}, \{V_{1}, V_{3}\}\}$, and $\{\{V_{3}\}, \{V_{1}, V_{2}\}\}$. In contrast, for two-dimensional systems, there is only one way to divide the vectors into two groups, and, therefore, only one function, ${u(x)}$, has to be constructed in order to perform BSS, as described in Subsection \ref{two-dimensional systems}. 

For each grouping, let $N_1$ and $N_2$ denote the number of vectors in groups $1$ and $2$, respectively, and let $G_1$ and $G_2$ denote the collections of values of $i$ for the vectors $V_{(i)}$ in groups $1$ and $2$, respectively. Each mapping, $u(x)$, is comprised of the union of the components of an $N_{1} \mbox{-component}$ function, $u_{(1)}(x)$, and the components of an $N_{2} \mbox{-component}$ function, $u_{(2)}(x)$, which are constructed as described in the next paragraph.  For example, for the above-mentioned three-dimensional system, the first mapping to be computed, $u(x)$, has three components, comprised of the single component of $u_{(1)}(x)$ and the two components of $u_{(2)}(x)$. 

The construction of $u_{(1)}(x)$ is initiated by picking any point $x_0$ in the $x$ coordinate system. We then find an $N_{1} \mbox{-dimensional}$ curvilinear subspace, consisting of all points that can be reached by starting at  $x_0$ and by moving along all linear combinations of the local vectors in group $1$. This subspace can be described by a function $X(\sigma)$, where the components of $\sigma$ ($\sigma_{i} \mbox{ for } i \in G_1$) parameterize the subspace by labelling its points in an invertible fashion. Formally, $X(\sigma)$ can be chosen to be a solution of the differential equations
\begin{equation}
\label{X ND}
\frac{\partial{X}}{\partial{\sigma_{i}}} = V_{(i)}(X)
\end{equation}
for $i \in G_1$ with the boundary condition, $X(0) = x_0$. Then, for each value of $\sigma$, we define an $N_{2} \mbox{-dimensional}$ curvilinear subspace, consisting of all points that can be reached by starting at $X(\sigma)$ and by moving along all linear combinations of the local vectors in group $2$. This subspace can be described by a function $Y(\tau)$, where the components of $\tau$ ($\tau_{j} \mbox{ for } j \in G_2$) parameterize the subspace by labelling its points in an invertible fashion. $Y(\tau)$ can be chosen to be a solution of the differential equations
\begin{equation}
\label{Y high dimension}
\frac{\partial{Y}}{\partial{\tau_{j}}} = V_{(j)}(Y)
\end{equation}
for $j \in G_2$ with the boundary condition, $Y(0) = X(\sigma)$. Finally, the function $u_{(1)}(x)$ is defined so that it is constant on each one of the $Y$ subspaces. Specifically, $u_{(1)}(x) \equiv \sigma$ whenever $x$ is in the $Y$ subspace containing $X(\sigma)$. The function $u_{(2)}(x)$ is defined by following an analogous procedure in which the roles of groups $1$ and $2$ are switched. Finally, the union of the $N_1$ components of $u_{(1)}(x)$ and the $N_2$ components of $u_{(2)}(x)$ is taken to define the mapping, $u(x)$, that corresponds to the chosen grouping of the vectors, $V_{(i)}$, into groups $1$ and $2$.

The foregoing procedure can be illustrated by considering the construction of $u_{(1)}(x)$ from the first grouping of vectors in the three-dimensional case mentioned in the previous paragraph. In that case: 
\begin{itemize}
\item[a)] $X(\sigma)$ describes a curved line that passes through $x_0$, that is parallel to $V_{(1)}(x)$ at each point, and that is parameterized by $\sigma$; 
\item[b)] each function, $Y(\tau)$, describes a curved surface, which intersects that curved line at some value of the parameter $\sigma$ and which is parallel to all linear combinations of $V_{(2)}(x)$ and $V_{(3)}(x)$ at each point;
\item[c)] along each of these curved surfaces, $u_{(1)}(x)$ is equal to the corresponding value of $\sigma$. 
\end{itemize}
Likewise, for the construction of $u_{(2)}(x)$ in the three-dimensional case: 
\begin{itemize}
\item[a)] $X(\sigma)$ describes a curved surface that passes through $x_0$, that is parallel to all linear combinations of $V_{(2)}(x)$ and $V_{(3)}(x)$ at each point, and that is parameterized by the two components of $\sigma$; 
\item[b)] each function $Y(\tau)$ describes a curved line, which intersects that surface at a value of the parameter $\sigma$ and which is parallel to $V_{(1)}(x)$ at each point; 
\item[c)] along each of these curved lines, $u_{(2)}(x)$ is equal to the corresponding value of $\sigma$.
\end{itemize}

\textit{3. Each mapping, $u(x)$, is used to transform the time series of measurements, $x(t)$, into a time series of transformed measurements, $u[x(t)]$.}\\
For each $u(x)$, the transformed time series, $u[x(t)]$, is the union of the $N_1$ components of $u_{(1)}[x(t)]$ and the $N_2$ components of $u_{(2)}[x(t)]$.\\

\textit{4. It is determined if at least one mapping leads to transformed measurements, $u[x(t)]$, having a density function that is the product of the density functions of $u_{(1)}[x(t)]$ and $u_{(2)}[x(t)]$.}\\
Specifically, it is determined if at least one transformed time series, $u[x(t)]$, has a PDF that factorizes as
\begin{equation}
\label{ND u PDF factorization}
\rho_{U}(u,\dot{u}) = \prod_{a=1,2}{\rho_{Ua}(u_{(a)},\dot{u}_{(a)})} .
\end{equation}
Here, $u$ denotes $u[x(t)]$, and $\dot{u}$ is its time derivative. Alternatively, we can compute a large set of correlations of multiple components of each transformed time series and then determine if they are products of lower-order correlations of two subsystems, as required by (\ref{ND u PDF factorization}).\\

\textit{5.  The result of step 4 is used to determine if the measurement data are separable and, if they are, to determine an unmixing function. Specifically, if at least one mapping, $u(x)$, produces a factorizable density function, it is obvious that the data are separable and $u(x)$ is an unmixing function. On the other hand, if none of the mappings leads to a factorizable density function, the data are inseparable in any coordinate system.}\\ 
This last statement is a consequence of the following fact, which is proved in the next two paragraphs: namely, if the data are separable, at least one of the mappings, $u(x)$, leads to a density function that is the product of the density functions corresponding to $u_{(1)}$ and $u_{(2)}$.

The only remaining task is to prove the above-mentioned consequence of separability. The first step is to show that the matrix $M$ and the local vectors have simple forms in the separable ($s$) coordinate system. In particular, we prove that the following block-diagonal matrix is the $M$ matrix in the $s$ coordinate system
\begin{equation}
\label{block-diagonal MS}
M_S(s) = \left( \begin{array}{ccc}  
   M_{S1}(s_{(1)}) & 0 &  \\
   0 & M_{S2}(s_{(2)}) &  .
   \end{array} \right) 
\end{equation}
Here, each submatrix $M_{Sa}$ is defined to be the $M$ matrix derived from the correlations between components of the corresponding subsystem state variable, $s_{(a)}$. For example, in the case of a separable three-dimensional system, (\ref{block-diagonal MS}) asserts that $M_S$ consists of $1 \times 1$ and $2 \times 2$ blocks, which are the $M$ matrices of one-dimensional and two-dimensional subsystems, respectively. In order to prove (\ref{block-diagonal MS}), it is necessary to show that $M_S$ satisfies (\ref{M definition 1}) and (\ref{M definition 2}) in the $s$ coordinate system. To do this, first note that (\ref{PDF moment}), (\ref{phase space factorization}), and (\ref{C_k=0}) imply that velocity correlations vanish in the $s$ coordinate system if their indices contain a solitary index from any one block. It follows that the second-order velocity correlation in the $s$ coordinate system ($C_{Skl}(s)$) consists of two blocks, each of which contains the second-order velocity correlations of an independent subsystem. This implies that (\ref{block-diagonal MS}) satisfies the constraint (\ref{M definition 1}), because, by definition, each block of $M_S$ transforms the corresponding block of $C_{Skl}$ into an identity matrix. In order to prove that (\ref{block-diagonal MS}) satisfies (\ref{M definition 2}), substitute it into the definition of
\begin{equation}
\label{ISklmm}
\sum_{1 \leq m \leq N} I_{Sklmm} .
\end{equation}
Then, note that: 1) when $k$ and $l$ belong to different blocks, each term in this sum vanishes because it factorizes into a product of correlations, one of which has a single index and, therefore, must vanish because of (\ref{C_k=0}); 2) when $k$ and $l$ belong to the same block and are unequal, each term with $m$ in any other block contains a factor equal to $I_{Skl}$, which vanishes for $k ~\neq~ l$, as proved above; 3) when $k$ and $l$ belong to the same block and are unequal, the sum over $m$ in the same block vanishes, because each block of $M_S$ is defined to satisfy (\ref{M definition 2}) for the corresponding subsystem. This completes the proof that $M_S$ satisfies (\ref{M definition 1}) and (\ref{M definition 2}). It follows that $M_S$ is the $M$ matrix in the $s$ coordinate system, as asserted above.

Recall that the local vectors in the $s$ coordinate system are columns of the matrix, $M^{-1}_S$. Because of the block diagonality of $M^{-1}_S$, the local vectors can be sorted into two groups (groups $1$ and $2$) that consist of the columns passing through blocks $1$ and $2$, respectively, and that contain $N_1$ and $N_2$ vectors, respectively. Therefore, at each point $s$, the local vectors in the first group are linear combinations of the unit vectors parallel to the first $N_1$ axes of the $s$ coordinate system, and the local vectors in the second group are linear combinations of the unit vectors parallel to the last $N_2$ axes of the $s$ coordinate system. Hence, in the $s$ coordinate system, the function, $X(\sigma)$, which was used to define $u_{(1)}(x)$, describes the linear subspace that contains the point $s[x_{0}]$ and that is spanned by the first group of unit vectors. Likewise, each $Y(\tau)$, which was used to define $u_{(1)}(x)$, describes a linear subspace that contains $s[X(\sigma)]$ for some value of $\sigma$ and that is spanned by the second group of unit vectors. This implies that the state variable of the first subsystem, $s_{(1)}(x)$, is constant within each $Y$ subspace, being equal to the value of $s_{(1)}(x)$ at the intersection of that $Y$ subspace with the $X$ subspace. But, recall that $u_{(1)}(x)$ is also constant within each $Y$ subspace, being equal to the value of $\sigma$ at its intersection with the $X$ subspace. Therefore, because $\sigma$ is defined to be invertibly related to the points in the $X$ subspace and because the values of $s_{(1)}$ are also invertibly related to the points in the $X$ subspace, these paired values must be invertibly related to one another; i.e., $\sigma = h_{1}(s_{(1)})$ where $h_1$ is an invertible function. It follows that $u_{(1)}(x)$ and $s_{(1)}(x)$ must also be invertibly related at each point; i.e., $u_{(1)}(x) = h_{1}[s_{(1)}(x)]$. In a similar manner, it can be shown that $u_{(2)}(x)$ and $s_{(2)}(x)$ are also related by some invertible function. Because $s_{(1)}$ and $s_{(2)}$ are the state variables of independent subsystems and because $u_{(1)}$ and $u_{(2)}$, respectively, are invertibly related to them, $u_{(1)}$ and $u_{(2)}$ must be subsystem state variables in some other subsystem coordinate systems. This completes the proof of the assertion at the beginning of the previous paragraph: namely, if the data are separable, at least one way of grouping the local vectors (e.g., the grouping corresponding to the above-mentioned blocks) leads to a mapping, $u(x)$, that describes a pair of statistically independent state variables ($u_{(1)}$ and $u_{(2)}$).

\section{Experiments}
\label{experiments}

\begin{figure}
\centering
\subfloat[]{\includegraphics[width=0.35\textwidth]{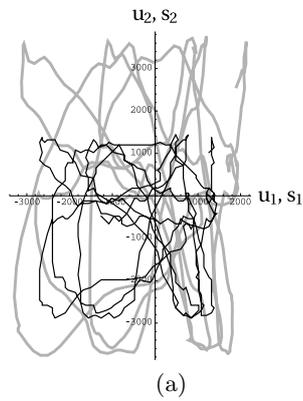}%
\label{fig_s(t)}}\par

\subfloat[]{\includegraphics[width=0.35\textwidth]{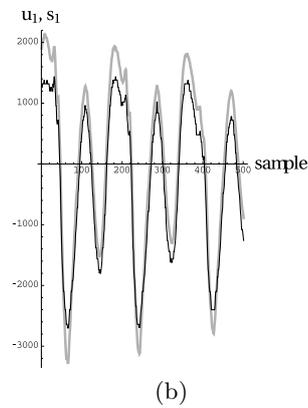}%
\label{fig_s1(t)}}\par

\subfloat[]{\includegraphics[width=0.35\textwidth]{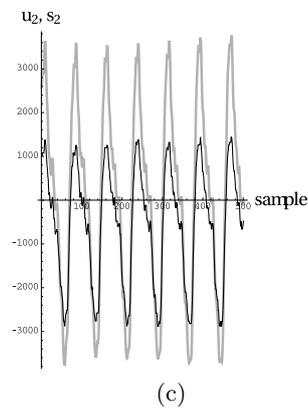}%
\label{fig_s2(t)}}

\caption{ (a) The thick gray line depicts the trajectory of 30 ms of the two speakers' speech in the $s$ coordinate system, in which each component is equal to one speaker's speech amplitude. The thin black line depicts the recovered waveforms ($u[x(t)]$) of the two speakers during the same time interval, computed by blindly processing their nonlinearly mixed speech.  Panels (b) and (c) show the time courses of $s_1$ and $u_1$ and of $s_2$ and $u_2$, respectively, during the same 30 ms time interval.}
\label{figure2}
\end{figure}

\begin{figure}
\centering
\subfloat[]{\includegraphics[width=0.35\textwidth]{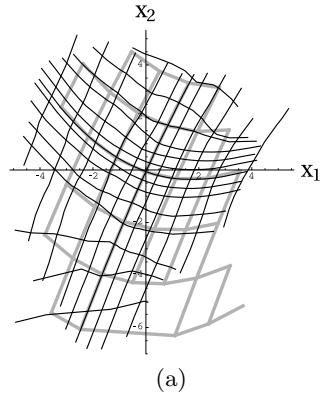}%
\label{fig_warped_grid}}\par

\subfloat[]{\includegraphics[width=0.35\textwidth]{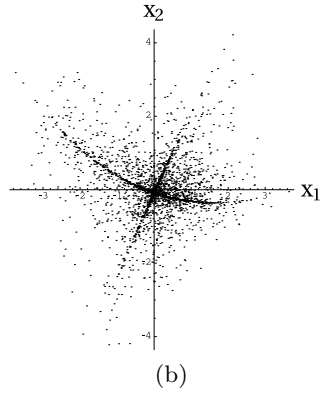}%
\label{fig_measurements}}\par

\subfloat[]{\includegraphics[width=0.35\textwidth]{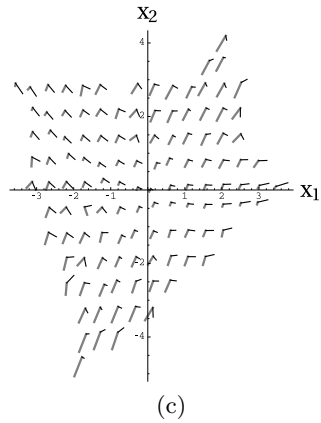}%
\label{fig_V}}

\caption{ (a) The thick gray curves comprise a regular Cartesian grid of lines in the $s$ coordinate system, after they were nonlinearly mapped into the $x$ coordinate system by the mixing in (\ref{mixing}). The thin black lines depict lines of constant $u_1$ or of constant $u_2$, where $u$ denotes a possibly separable coordinate system derived from the measurements, $x(t)$. (b) A random subset of the measurements along the trajectory of the mixed waveforms, $x(t)$. (c) The thick gray and thin black lines show the local vectors, $V_{(1)}$ and $V_{(2)}$, respectively, after they have been uniformly scaled for the purpose of display.}
\label{figure3}
\end{figure}

\begin{figure}
\centering
\subfloat[]{\includegraphics[width=0.35\textwidth]{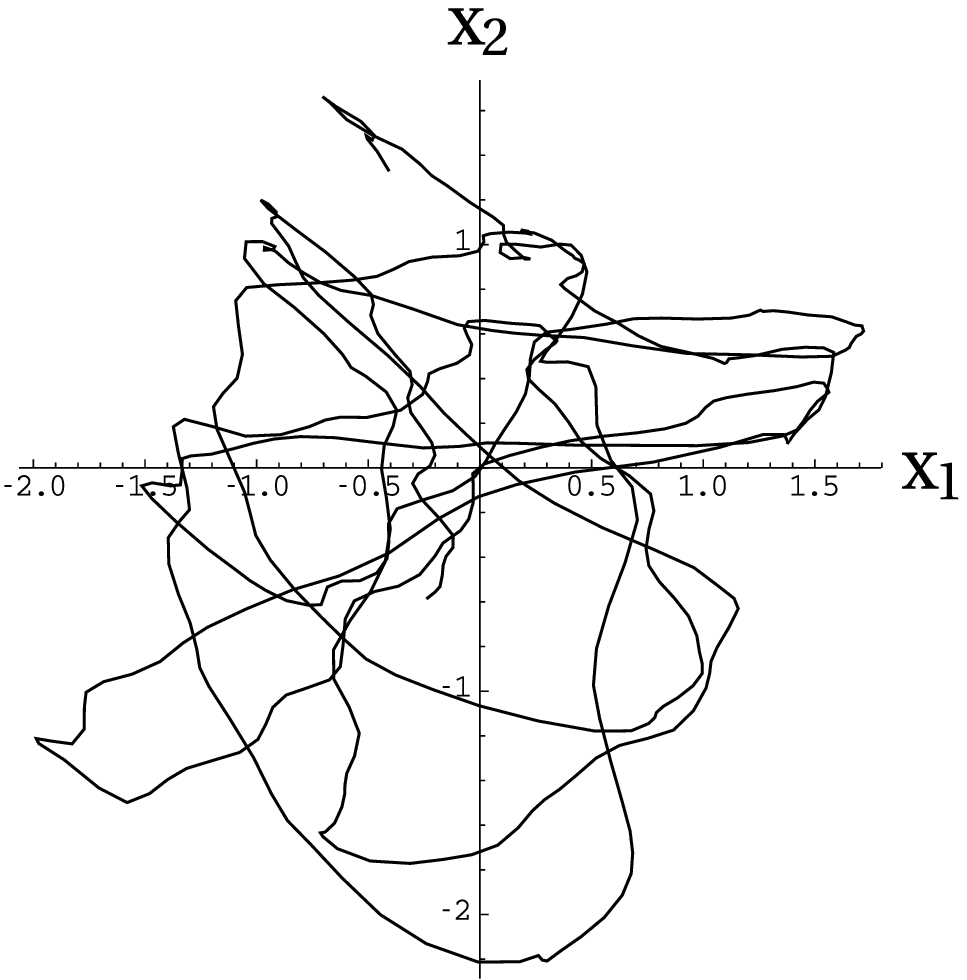}%
\label{fig_x(t)}}\par

\subfloat[]{\includegraphics[width=0.35\textwidth]{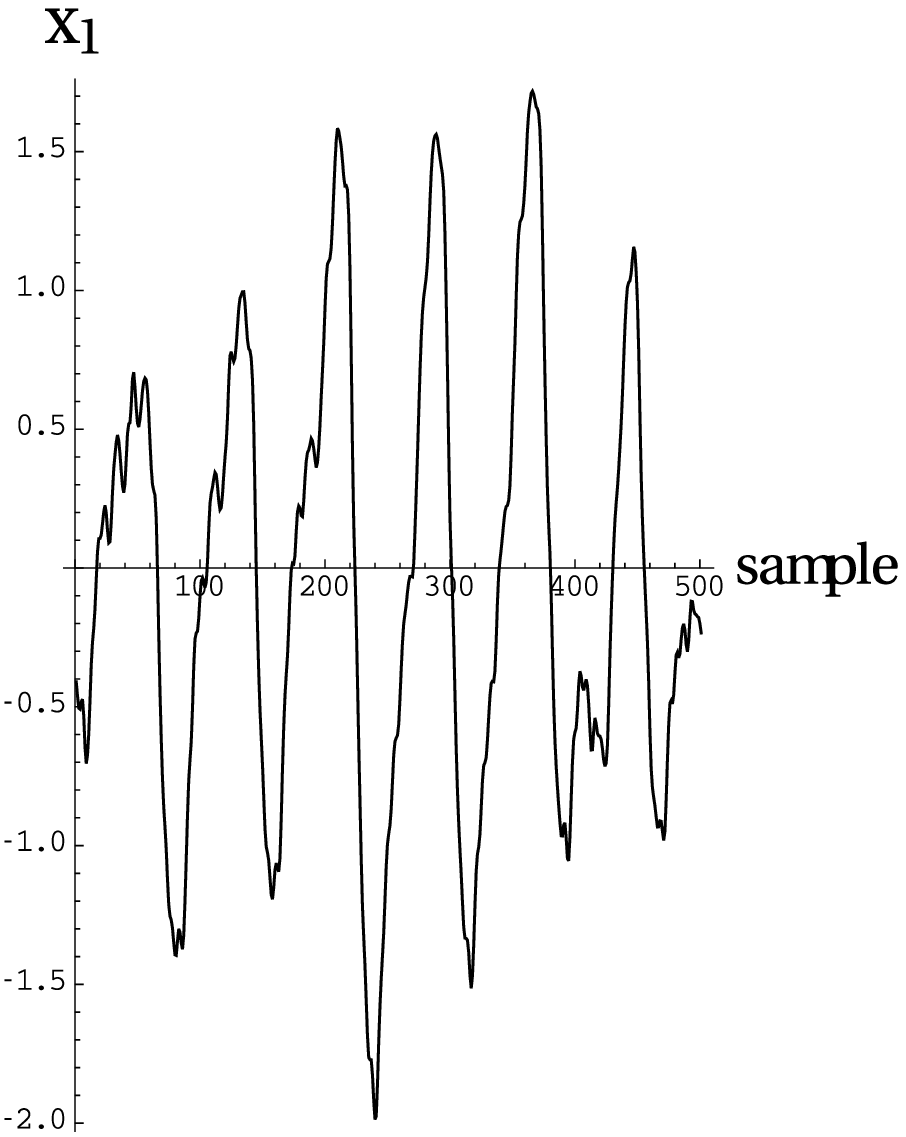}%
\label{fig_x1(t)}}\par

\subfloat[]{\includegraphics[width=0.35\textwidth]{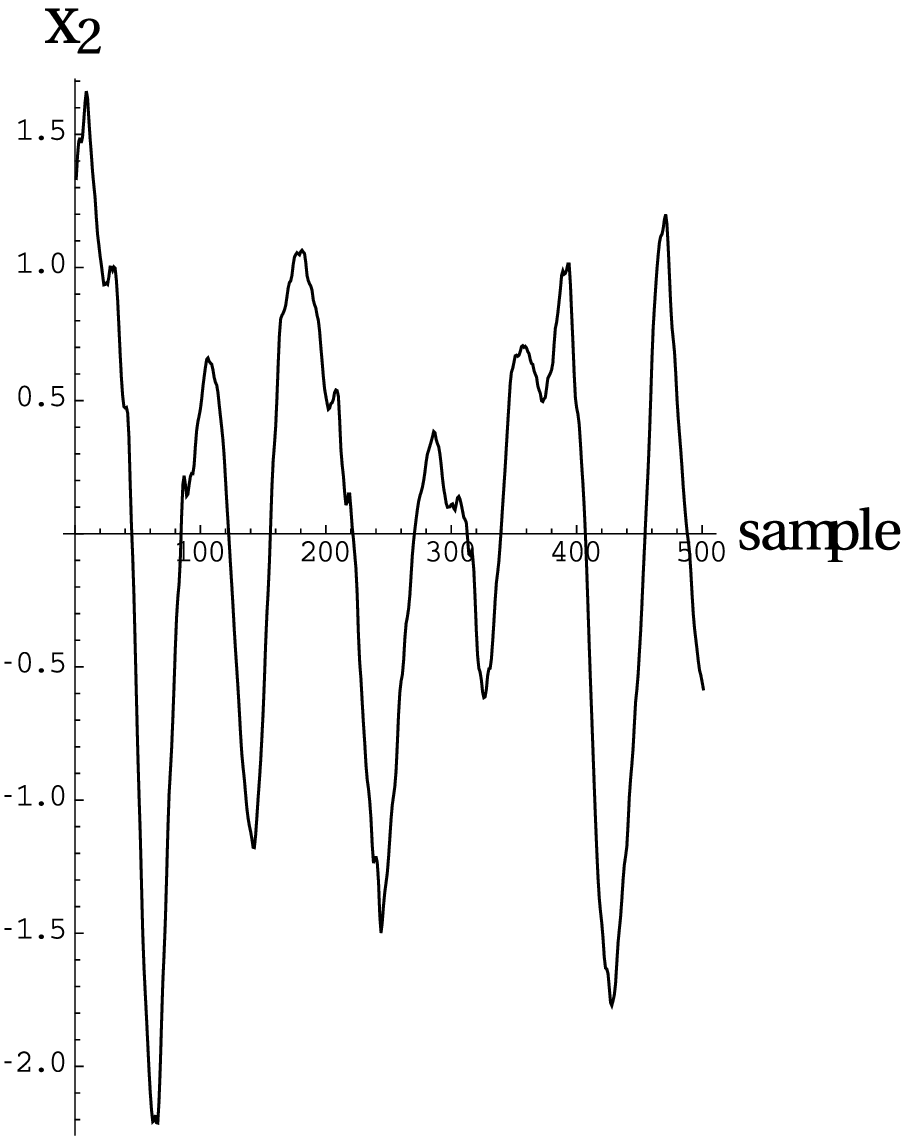}%
\label{fig_x2(t)}}

\caption{(a) The trajectory of measurements, $x(t)$, during the 30 ms time interval depicted in Figure \ref{figure2}.  Panels (b) and (c) show the time courses of $x_1$ and $x_2$, respectively, during the same 30 ms time interval.}
\label{figure4}
\end{figure}

In this section, the new BSS technique is illustrated by using it to disentangle synthetic nonlinear mixtures of two audio waveforms. The audio waveforms consisted of two thirty-second excerpts from audio books, each one read by a different male speaker. The waveform of each speaker, denoted $s_{k}(t)$ for $k=1 \mbox{ or } 2$, was sampled 16,000 times per second with two bytes of depth. The thick gray lines in Figure \ref{figure2} show the two speakers' waveforms during a short (30 ms) interval. These waveforms were then  mixed by the nonlinear functions
\begin{equation}
\label{mixing}
\begin{split}
\mu_{1}(s) &= 0.763 s_1 + (958 - 0.0225 s_2)^{1.5} \\
\mu_{2}(s) &= 0.153 s_2 + (3.75 * 10^7-763 s_1 - 229 s_2)^{0.5} ,
\end{split}
\end{equation}
where $-2^{15} \leq s_1, s_2 \leq 2^{15}$. This is one of a variety of nonlinear transformations that were tried with similar results. The synthetic mixture measurements, $x_{k}(t)$, were taken to be the variance-normalized, principal components of the sampled waveform mixtures, $\mu_{k}[s(t)]$. Figure \ref{fig_warped_grid} shows how this nonlinear mixing mapped an evenly-spaced Cartesian grid in the $s$ coordinate system onto a warped grid in the $x$ coordinate system. Figure \ref{fig_measurements} shows the distribution of the synthetic measurements created by randomly sampling $x(t)$, and Figure \ref{figure4} shows the time course of $x(t)$ during the same short time interval depicted in Figure \ref{figure2}. When either waveform mixture ($x_{1}(t)$ or $x_{2}(t)$) was played as an audio file, it sounded like a confusing superposition of two voices, which were quite difficult to understand. 

The proposed BSS technique was then applied to these synthetic measurements as follows:
\begin{enumerate}
\item The entire set of 500,000 measurements, consisting of $x$ and $\dot{x}$ at each sampled time, was sorted into a $16 \, \times \, 16$ array of bins in $x \mbox{-space}$. Then, the $\dot{x}$ distribution in each bin was used to compute local velocity correlations (see (\ref{C2 definition}) and (\ref{C4 definition})), and these were used to derive $M$ and $V_{(i)}$ for each bin. Figure \ref{fig_V} shows these local vectors at each point. 
\item These vectors were used to construct the mapping, $u(x)$. As described in Method, the first step was to choose some point $x_0$ and then construct the curvilinear line, $X(\sigma)$, that passes through that point and is tangential to the local vector $V_{(1)}(x)$ everywhere. Then, at each point $\sigma$ on this curve, a curvilinear line, $Y(\tau)$, was constructed through it so that it was tangential to the local vector $V_{(2)}(x)$ everywhere. 
Along each of these $Y$ curves, $u_{1}(x)$ was defined to be a constant equal to the value of $\sigma$ at the curve's point of intersection with $X(\sigma)$. The mapping, $u_{2}(x)$, was defined by an analogous procedure. In this way, each point $x$ was assigned values of both $u_{1}$ and $u_{2}$, thereby defining the mapping, $u(x)$. One of the groups of thin black lines in Figure \ref{fig_warped_grid} depicts a family of curves having constant values of $u_1$, which are evenly-spaced and increase as one moves from curve to curve in the family. The other group of thin black lines in Figure \ref{fig_warped_grid} shows a family of curves having constant values of $u_2$, which are evenly-spaced and increase as one moves from curve to curve in the family. 
\item As proved in Method, if the data are separable, $u(x)$ must an unmixing function. Therefore, the separability of the data could be determined by seeing if $u[x(t)]$ has a factorizable density function (or factorizable correlation functions). If the density function does factorize, the data are patently separable, and $u_{1}[x(t)]$ and $u_{2}[x(t)]$ describe the evolution of the independent subsystems. On the other hand, if the density function does not factorize, the data must be inseparable.
\end{enumerate}  
In this illustrative example, the separability of the $u$ coordinate system was verified by a more direct method. Specifically, Figure \ref{fig_warped_grid} shows that the isoclines for increasing values of $u_1$ (or $u_2$) nearly coincide with the isoclines for increasing values of $s_1$ (or $s_2$). This demonstrates that the $u$ and $s$ coordinate systems differ by component-wise transformations of the form: $(u_{1}, u_{2}) = (h_{1}(s_{1}), h_{2}(s_{2}))$ where $h_1$ and $h_2$ are monotonic functions.  Because the data are separable in the $s$ coordinate system and because component-wise transformations do not affect separability, the data's PDF must factorize in the $u$ coordinate system. Therefore, we have accomplished the objectives of BSS: namely, by blindly processing the mixture measurements, $x(t)$, we have determined that the system is separable, and we have computed the transformation, $u(x)$, to a separable coordinate system.

The transformation, $u(x)$, can be applied to the mixture measurements, $x(t)$, to recover the original unmixed waveforms, up to component-wise transformations.  The resulting waveforms, $u_{1}[x(t)]$ and $u_{2}[x(t)]$, are depicted by the thin black lines in Figure \ref{figure2}, which also shows the trajectory of the unmixed waveforms in the $s$ coordinate system. Notice that the two trajectories, $u[x(t)]$ and $s(t)$, are similar except for component-wise transformations along the two axes. A component-wise transformation is especially noticeable as a stretching of $s_{2}(t)$ with respect to $u_{2}[x(t)]$ along the positive $s_2$ axis. When each of the recovered waveforms, $u_{1}[x(t)]$ and $u_{2}[x(t)]$, was played as an audio file, it sounded like a completely intelligible recording of one of the speakers. In each case, the other speaker was not heard, except for a faint ``buzzing'' sound in the background. Therefore, the component-wise transformations (e.g., the above-mentioned ``stretching''), which related the recovered waveforms to the original unmixed waveforms, did not noticeably reduce intelligibility. 

\section{Conclusion}

This paper describes how to determine the separability of time-dependent measurements of a system, $x(t)$; namely, it shows how to determine if there is a linear or nonlinear function (an unmixing function) that transforms the data into a collection of signals from statistically independent subsystems. First, the measurement time series is shown to endow state space with a local structure, consisting of vectors at each point $x$. If the data are separable, each of these vectors is directed along a subspace traversed by varying the state variable of one subsystem, while all other subsystems are kept constant. Because of this property, these vectors can be used to derive a small number of mappings, $\{u(x)\}$, which must include an unmixing function, if one exists. In other words, the data are separable if and only if one of the $u(x)$ describes a separable coordinate system. Therefore, separability can be determined by testing the separability of the data, after they have been transformed by each of these mappings.

Some comments on this result:
\begin{enumerate}
\item The original problem of looking for an unmixing function, $f(x)$, among an \emph{infinite set of functions} was reduced to the simpler problem of constructing a \emph{small number of mappings}, $\{u(x)\}$, and then determining if one of them transforms the data into separable form.

\item The BSS method described in this paper is model-independent in the sense that it can be used to separate data that were mixed by any invertible diffeomorphic mixing function. In contrast, most other approaches to nonlinear BSS are model-dependent because they assume that the mixing function has a specific parametric form (\cite{Duarte},\cite{Merrikh},\cite{Taleb}).

\item Notice that the proposed method is analytic and constructive in the sense that the candidate unmixing functions are constructed directly from the data, by locally manipulating them with linear algebraic techniques. In contrast, many other approaches \cite{Comon Jutten} search for an unmixing function by utilizing more complex techniques, involving neural networks or iterative computations.

\item Theoretically, the proposed method can be applied to measurements described by any diffeomorphic mixing function. However, more data will have to be analyzed in order to handle mixing functions with more pronounced nonlinearities. This is because rapidly varying mixing functions may cause the local vectors ($V_{(i)}$) to vary rapidly in the measurement coordinate system, making it necessary to compute those vectors in numerous small neighborhoods.

\item More data will also be required to apply this method to systems with many degrees of freedom. In Experiments, thirty seconds of data (500,000 samples) were used to recover two audio waveforms from measurements of two nonlinear mixtures. In other experiments, approximately six minutes of data (6,000,000 samples) were used to cleanly recover the waveforms of four sound sources (two speakers and two piano performances) from four signal mixtures. As expected, blind separation for the 4D state space did require more data, but it was not a prohibitive amount.

\item The proposed method does not require unusual computational resources. In any event, the most computationally expensive tasks are the binning of the measurement data and the computation of the local vectors, $V_{(i)}$, in each bin. If necessary, these calculations can be parallelized across multiple CPUs.

\item This paper shows how to perform nonlinear BSS for the case in which the mixture measurements are invertibly related to the state variables of the underlying system. Invertibility can almost be guaranteed by observing the system with a sufficiently large number of independent sensors: specifically, by utilizing at least $2N+1$ independent sensors, where $N$ is the dimension of the system's state space. In this case, the sensors' output lies in an $N \mbox{-dimensional}$ subspace embedded within a space of at least $2N+1$ dimensions. Dimensional reduction techniques (e.g., \cite{Roweis}) can be used to find the subspace coordinates corresponding to the sensor outputs.  Because an embedding theorem asserts that this subspace is very unlikely to self-intersect \cite{Sauer}, the coordinates on this subspace are almost certainly invertibly related to the system's state space, as desired.

\item Separability is an intrinsic or coordinate-system-independent property of data; i.e., if it is true (or false) in one coordinate system, it is true (or false) in all coordinate systems. The local vectors ($V_{(i)}$) also represent a kind of intrinsic structure on state space, and, as mentioned previously, these contain some information about separability, which is available in all coordinate systems. These vectors ``mark'' state space and are analogous to directional arrows, which mark a physical surface and which can be used as navigational aids, no matter what coordinate system is being used. Many other vectors can be derived from the local velocity distributions of a time series. However, most of them will not have the special property of the $V_{(i)}$: namely, the property of being aligned with the directions traversed by the system when just one subsystem is varied and all others are held constant. For example, the $V_{(i)}$ would not have this critical property if the definition of $M$ (see (\ref{M definition 1}) and (\ref{M definition 2})) was changed by replacing $\sum_{1 \leq m \leq N} I_{klmm}$ with higher order correlations (e.g., $\sum_{1 \leq m, \, n \leq N} I_{klmmnn}$).
\end{enumerate}

%\section{References}
\bibliographystyle{IEEEtran}

\end{document}